\documentclass{egpubl-eurovis-short}
\usepackage{eurovis2014-short}

\WsPaper         

\electronicVersion 

\ifpdf \usepackage[pdftex]{graphicx} \pdfcompresslevel=9
\else \usepackage[dvips]{graphicx} \fi

\PrintedOrElectronic

\usepackage{t1enc,dfadobe}

\usepackage{egweblnk}
\usepackage{cite}

\title{Interactive Visual Exploration of Topic Models using Graphs}

\author[Samuel R\"onnqvist, Xiaolu Wang \& Peter Sarlin]
       {Samuel R\"onnqvist$^{1,2}$,\ \ Xiaolu Wang$^{2}$\ \ \&\ \ Peter Sarlin$^{3,4}$
        \\
         $^1$Turku Centre for Computer Science \ \ 
         $^2$\AA bo Akademi University, Finland \ \
         $^3$Goethe University, Center of Excellence SAFE, Germany \\
         $^4$RiskLab at IAMSR, \AA bo Akademi University and Arcada University of Applied Sciences
       }

\begin{document}

\maketitle

\begin{abstract}

Probabilistic topic modeling is a popular and powerful family of tools
for uncovering thematic structure in large sets of unstructured text
documents. While much attention has been directed towards the modeling
algorithms and their various extensions, comparatively few studies
have concerned how to present or visualize topic models
in meaningful ways. In this paper, we present a novel design that
uses graphs to visually communicate topic structure and meaning. By
connecting topic nodes via descriptive keyterms, the graph representation
reveals topic similarities, topic meaning and shared, ambiguous keyterms.
At the same time, the graph can be used for information retrieval
purposes, to find documents by topic or topic subsets. To exemplify
the utility of the design, we illustrate its use for organizing and
exploring corpora of financial patents.


\end{abstract}

\section{Introduction}

Across domains, we are faced with substantial and often overwhelming
amounts of textual data, which present a need for tools to aid in
organizing and understanding its content. Text mining techniques are
widely used to computationally model human language and extract meaning.
Granted, text mining has been successfully applied in many areas,
yet, the intricacies of human language constitute an ever-present
challenge to computational processing of text. Human involvement is
still central to the text mining process. While computational modeling
can scan vast data and extract information that is likely to be interesting,
humans are uniquely capable of a deeper understanding\cite{risch2008text}. Acknowledging
the essential roles of both computational and human information processing
in text mining, the necessity for effective communication between
the two parts becomes apparent. Visualization is a highly efficient
communication channel, and the two-way communication enabled by interaction supports a more intimate understanding of the underlying model \cite{Ware2004}.

This paper is concerned with probabilistic topic modeling and visual
communication of modeling results. Topic modeling algorithms are used
to discover latent topics in sets of documents, as a way of uncovering
thematic structure. However, visualization of topic models is still
a little researched area, which we seek to explore. This answers directly
to concerns voiced by one of the authors of the original probabilistic
topic modeling algorithm, David M. Blei, who states that ``topic models
provide new exploratory structure in large collections -- how can
we best exploit that structure to aid in discovery and exploration?''\cite{blei2012probabilistic}.
He further asserts that interface questions are ``essential to topic
modeling'', but that ``making this structure useful requires careful
attention to information visualization and {[}...{]} user interfaces''.

Our contribution is a new visualization design that aims to convey
the meaning of found topics in an intuitive way, through a \emph{topic cloud}, as well as how topics
relate to each other. It is an interactive graph visualization that
connects topics and descriptive keyterms, providing both corpus overview
and thematic exploration of documents. A web-based implementation is available at: http://risklab.fi/demo/topics/. We illustrate the use of this
visualization design through a case, in which topics in financial
patent applications are modeled and presented for exploration to support domain experts
in satisfying their information needs. The lack of proper tools to organize
patents has come to be a pressing problem \cite{elman2007automated}
that we seek to alleviate through the combination of computational processing and visual
interfaces.

\section{Probabilistic Topic Modeling}

The family of probabilistic topic modeling algorithms has emerged
as a widely popular approach to analyzing thematic structure in text.
Latent Dirichlet Allocation (LDA) is the most fundamental
among them, having inspired a long array of variations\cite{blei2009topic}.
Topic modeling is applicable especially to problems where there is
little prior knowledge of the content of the text, i.e.,
where exploratory analysis is needed. LDA infers latent topics in an unsupervised manner, based on word
co-occurrence in documents, and provides interpretable output in the
form of probabilities. The algorithm takes the desired number of topics
as input, assumes that each document may discuss several topics and
attempts to identify coherent and meaningful topics by analyzing the
terms in each document. By taking the context into account, LDA can
help disambiguate single words.

There are two types of relations that are interesting for the presentation
of LDA results: the topic-document relations and the topic-term relations.
First, LDA provides a probability distribution over topics for each
document, that is, to what degree a document relates to each of the
topics. In the scope of this paper, this topic-document information
is useful for retrieving documents based on topics of interest. Second,
LDA provides topic assignments for each term in a document. The topics
provided by LDA are defined by probability distributions over terms.
The elementary way of representing the meaning of a topic is through
its term probabilities directly, which are based on term frequencies
for the topic. Such a representation is rather uninformative to a
user, as it gives high ranking to common stop words (e.g., the, a,
and) and terms that are general to the whole document corpus (in the
case of patents: method, system etc.); better solutions
for topic presentation are discussed in Section 3.1.

\begin{figure*}
\centering
\includegraphics[width=1\textwidth]{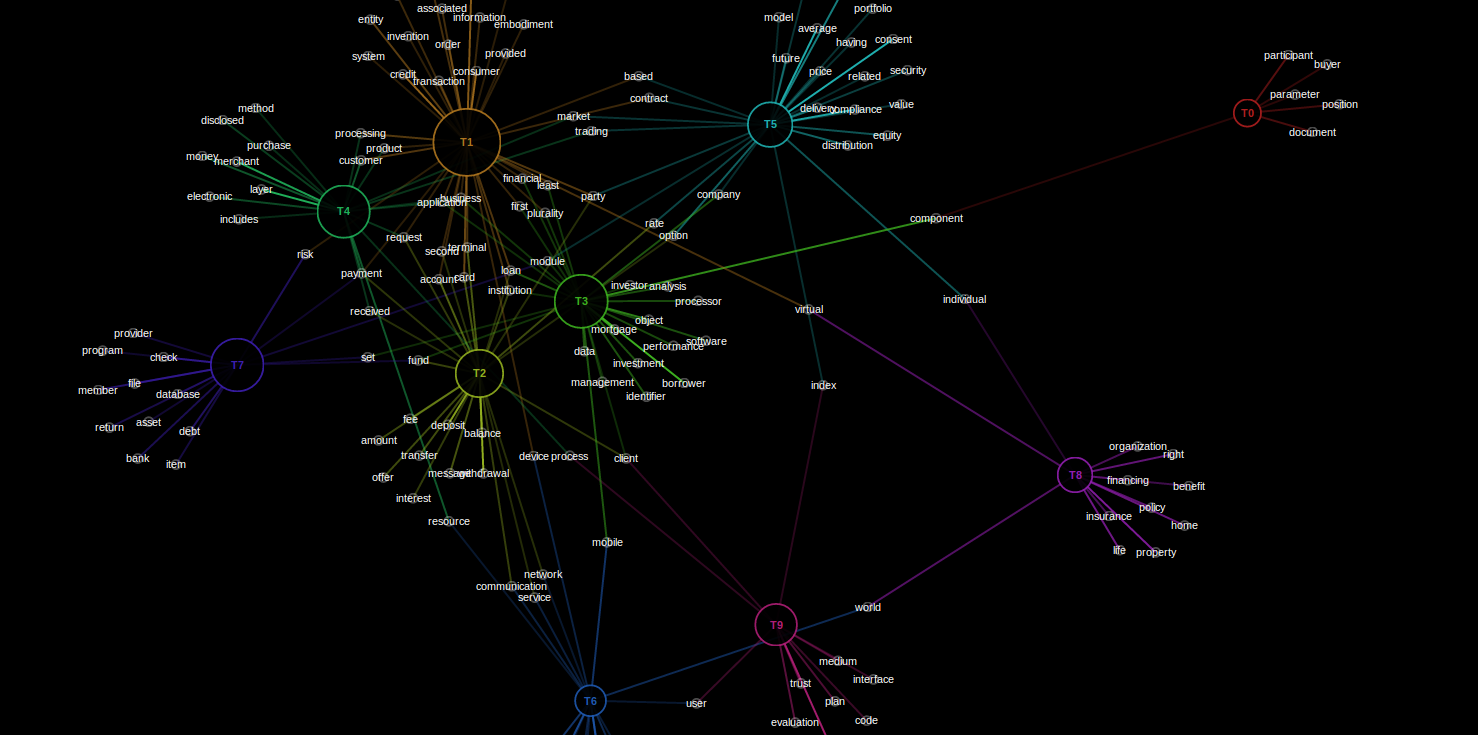}\caption{termvOverview of all topics. Topics that share keyterms are linked and reside closer,
link strength represents how distinguishing a keyterm is of a topic,
topic node size represents prevalence in the corpus.}
\end{figure*}

\section{Visualization of Topic Models}

Despite the importance carried by visual representation in making
topic models useful, the subject
has seen limited research effort. Some tools that present topic modeling
results rely heavily on text ordered in different fields, but use
few visual aids to communicate structure (see, e.g., \cite{chaney2012visualizing,gardner2010topic}).
A few other notable tools can be found that use visualization more
extensively, such as those by Chuang et al. \cite{chuang2012termite}
and Gretarsson et al. \cite{gretarsson2012topicnets}. Still, much
room is left for further exploration of how visualization techniques
can improve communication of topic modeling results, and results from
text mining models in general. This exploration is also supported
by Tufte's \cite{Tufte1983} design principle, to use graphics when
words alone cannot communicate the message effectively, as structural
information is as central to the analysis of text as semantics.

Our work extends that of Chuang et al. \cite{chuang2012termite},
which we find to be the most promising previous example of topic model
visualization. They visualize topic-keyterm relations through a matrix
view, which provides some idea of topic distribution, similarities
and meaning. Our approach may be seen as a graph visualization that
uses a topic-keyterm matrix similar to theirs as an adjacency matrix.
We argue that the force-directed graph visualization provides a view
that more intuitively communicates topic similarity structure. In the following, we discuss the three main components of the interface:
keyterm selection, graph representation and retrieval of documents.

\subsection{Selecting informative topic keyterms}

The raw topic-term probabilities provided by LDA are not suitable
as a ranking to find terms that distinguish well between topics, as
previously mentioned. Various measures have been proposed for reranking
that provide better description of the topic meaning, such as a TFIDF-inspired
\emph{term-score}\cite{blei2009topic} and others that likewise penalize
terms that are prevalent in many topics. Reranking the terms is essential
to produce descriptive and distinguishing keyterms for presentation
and visualization of topic meaning. 

For the sake of interpretability, we use the conditional probability
$P(T|w)$ to score how distinguishing a term is of its topic. Given
that a term \emph{w} is observed, the probability of it belonging
to topic \emph{T} is a measure of how distinguishing \emph{w} is of
\emph{T}. We derive the measure as $P(T|w)=P(w|T)P(T)/P(w)$ where
$P(w|T)$ is the topic-term probability distribution provided by LDA.
The use of $P(T|w)$ to identify informative terms is in line with
\cite{chuang2012termite}. A threshold on the probability selects the top keyterms for each topic. Additionally, a limit on number of keyterms per topic may
be enforced to avoid cluttering the visualization, or shared keyterms
may be prioritized to shift focus towards topic relations rather than
individual description.

\subsection{Visualizing topics and keyterms as a graph}

As our basic visualization technique, we use a graph with force-directed layout (using the D3 force algorithm\cite{bostock2011d3}). It provides a spatial metaphor
for topic similarity in the corpus. The graph, as shown in Figure 1, consists of
topic nodes connected to keyterm nodes only, which produces a fairly
sparse graph for which force-directed layouting can produce clear
results. Topics are not directly connected to each other, rather only
through their common keyterms. Still, the node positioning communicates
general topic similarity structures, and the connecting keyterms provide
qualitative detail on the nature of their relation. Visual connectedness
is a strong means for communicating relationships \cite{palmer1994rethinking}. The figure shows a (partial) overview of topics, where:
\begin{itemize}
\item the keyterm weighting is encoded by link opacity to communicate the strength of the relation 
\item node size is relative to the general frequency of the topic
\item colors from a qualitative scale are used to better distinguish the topic neighborhoods.
\end{itemize} 
Nodes can be dragged to alter their positions, which are then adjusted
by the force-directed algorithm in real time. However, the automatic
adjustment is slow enough to allow the user to move several nodes
before a stable conformation is reached, which allows for interactive
exploration of alternative locally optimal conformations and helps
the user inspect the structure of the graph. Our implementation also
allows zooming and panning to support exploration of many topics even
on smaller screens.

The meaning of a topic is represented by the keyterms linked to it,
each with the weighting discussed in the previous section. While the
initial view of Figure 1 provides an overview of all topics, focusing
on a single topic by hovering highlights its details, as in Figure 2.
Highlighting fades all parts of the graph not connected
to the topic to preattentively direct the user and ease their inspection,
in other terms, it provides context plus focus\cite{card1999readings}.
The topic-specific term weighting can now also be encoded through
keyterm font size, which enables simultaneous reading of term meaning
and importance. Similar to a tag cloud, this creates an easy-to-read
\emph{topic cloud}. 

Some terms are shared among topics, which hints at their ambiguity.
Ideally, a topic represents a meaningful context in which a term is
used, through which the sense of the term is made clear. Hence, two
topics sharing a term, as seen in Figure 3, might indicate that the term holds two different
meanings in the corpus. Handling such ambiguity is a central purpose
of topic models, but simple topic representations do not facilitate
identification of such patterns by the user, while visualization easily
can.

\subsection{Finding documents by topic and keyterm}

As the graph displays only topic nodes and a limited set of keyterm
nodes, but no document information, it acts as an abstracted topical
view of the corpus. All document information is accessible
only through interaction, which means that scaling the corpus size
does not affect the visualization.

By double clicking on nodes, any number and type of nodes can be selected
to function as anchors for the underlying documents. This mechanism
is used to drill down on the documents related to interesting topics,
keyterms or their combinations. The titles of the documents are listed
in a side panel linking to the full texts. Both focusing on nodes
and selecting them follow the visual information seeking mantra\cite{Shneiderman1996},
by filtering information and providing details on demand, about topics,
keyterms or related documents. Selecting a topic node ranks documents linked to it according to LDA's
topic-document distribution. Selecting a keyterm node of \emph{w }ranks
documents \emph{d }according to $P(d|w)$. The documents are ranked according to the joint probability when multiple nodes are selected

\begin{figure}
\centering
\includegraphics[height=8.2cm]{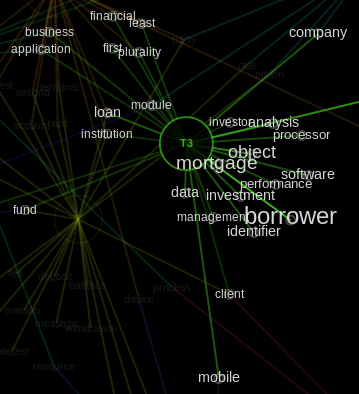}\caption{Focus on single topic neighborhood. Interpretation is aided by highlighting and weighting of keyterms.}
\end{figure}

\section{Case: Visualizing Topics in Financial Patents }

An example of abundant text data causing problems are patents. In particular, financial patents
(or business method patents) have witnessed an explosion
in numbers, referred to as the \emph{patent
flood} \cite{meurer2002business}, which has resulted from
a decline of business method exception to patentability due to court decisions. It has led to the issuance of low-quality patents, which further obstructs effective prior art search, creating uncertainty among inventors or would-be commercializers
of innovations \cite{hall2009business}. The situation has directed
considerable research effort in information retrieval towards patent
search\cite{tait2011current}, which still, nevertheless, relies on
visual analysis only to a very limited extent.

We demonstrate how topic modeling and our proposed visualization design
could provide effective means for organizing and visualizing financial
patents, by analysing a sample of 3954 abstracts of patent applications
filed between 2001 and 2011. The patents are classified by the U.S.
Patent and Trademark Office (USPTO) under patent subclass 705/35 defined
as finance (e.g., banking, investment or credit). As a basic example, we construct a topic model consisting of 10 topics,
for which an overview is shown in Figure 1. It is immediately clear
that topic T1 is the most frequent in the corpus (largest node size)
and that it is thematically central to the corpus (well connected
to other topics and positioned near the center). The user can visually browse the structural properties of the model, and compare the prevalence, centrality and semantics of topics. While the keyterms function as initial guidance in browsing, they can be challenging
to read in the more densely populated areas of the graph, and a focused
view is required to ease their inspection. Figure 2 demonstrates this
focused view on topic T3, with the rest of the graph faded in the
background as visual context. It is now easy to interpret that T3
represents patents about loan serivices by reading its highlighted keyterms,
weighted by how distinguishing they are of T3. For instance, we also identify T6 to discuss authentication systems. It is the task of the domain expert
to infer meaning and higher-level descriptions for the topics, to
further differentiate them and decide what best answers to their information
needs. Topic modeling and our presentation is helpful as it can provide a more
granular organization than the USPTO classifications.

The graph can be used to further explore the corpus, at the more detailed
level of keyterms. As Figure 3 illustrates, the term \emph{mobile}\textbf{\emph{
}}connected to T3 can be highlighted, upon which its connection to
T6 is easier to identify. This view communicates that \emph{mobile}
is used in two senses in the corpus: in the contexts of loan services
and authentication systems. This offers the user the possibility to
disambiguate the term in search for related documents. By selecting
T3 and \emph{mobile}, patents mentioning \emph{mobile} in the sense
of T3 are found, ranked and presented to the user. This mechanism
turns the visualization into a tool for information retrieval, not
only for high-level browsing of the corpus.

\begin{figure}
\centering
\includegraphics[height=8.2cm]{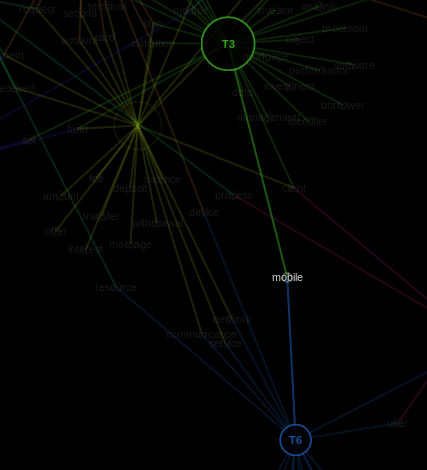}\caption{Focus on shared keyterm. Hovering over a shared keyterm highlights its topical contexts where it is used.}
\end{figure}

\section{Conclusions}

We have discussed a new visualization design for presenting LDA models
using graphs, in terms of topic structure and meaning, while retaining
a certain level of abstraction. Its use is demonstrated by modeling the topics in a set of financial patent abstracts, where it can aid information search problems. We focused explicitly on model exploration, assuming
the modeling results are satisfactory, although, real-world use of
topic models requires assessment of model quality by domain experts.
Continued studies of this type of visualization should consider that
topic modeling results are seldom perfect and the implications that
might have on interface design.

\bibliographystyle{eg-alpha}
\bibliography{references}

\end{document}